\documentclass[%
reprint,
showpacs,
 amsmath,amssymb,
 aps,
prl,
superscriptaddress,
preprintnumbers, showkeys]{revtex4-1}
\usepackage{textgreek}
\usepackage{color}
\usepackage{graphicx}
\usepackage{hyperref}

 \usepackage{xassoccnt}
 \newcounter{totalequations}
 \DeclareAssociatedCounters{equation}{totalequations}
 \newcounter{totalfigures}
\DeclareAssociatedCounters{figure}{totalfigures}

\hypersetup{bookmarksnumbered, pdfpagemode=UseOutlines, 
colorlinks=true, citecolor=blue, filecolor=blue, linkcolor=blue, urlcolor=blue}

 \let\theOldHfigure\theHfigure
 \renewcommand{\theHfigure}{\theOldHfigure::\number\value{totalfigures}}
 \let\theOldHequation\theHequation
 \renewcommand{\theHequation}{\theOldHequation::\number\value{totalequations}}

  \newcommand{\beginsupplement}{%
          \setcounter{table}{0}
          \renewcommand{\thetable}{S\arabic{table}}%
          \setcounter{figure}{0}
        \renewcommand{\thefigure}{S\arabic{figure}}
          \setcounter{equation}{0}
          \renewcommand{\theequation}{S\arabic{equation}}%
      }
  \makeatletter
  \let\cat@comma@active\@empty
  \makeatother

\begin{document}


\title{A two-channel model for Spin-relaxation noise}

\author{S. Omar} 
\email{s.omar@rug.nl}
\author{B.J. van Wees}
\affiliation{The Zernike Institute for Advanced Materials, University of Groningen Nijenborgh 4 9747 AG, Groningen, The Netherlands}%

\author{I.J. Vera-Marun}
\email{ivan.veramarun@manchester.ac.uk}
\affiliation{School of Physics and Astronomy, The University of Manchester, Manchester M13 9PL, UK}
\date{\today}

\begin{abstract}
 We develop a two-channel resistor model for simulating spin transport with general applicability. Using  this model, for the case  of graphene as a prototypical material, we calculate the spin signal consistent with experimental values. Using the same model we also simulate the charge and spin-dependent $1/f$ noise, both in the local and nonlocal four-probe measurement schemes, and identify the noise from the spin-relaxation resistances as the major source of spin-dependent $1/f$ noise.
\end{abstract}

     
\maketitle


Signal fluctuations with $1/f$ power spectral density are believed to originate from a broad distribution of time scales related with the measured quantity  \cite{dutta_low-frequency_1981, balandin_low-frequency_2013}, which for electronic transport is associated to the trapping-detrapping times of charge carriers via impurities \cite{dutta_low-frequency_1981, pal_microscopic_2011}. 
A two dimensional sheet of graphene, owing to its surface sensitivity \cite{fabian_relaxation_sub, Folk_relaxation, fabian_resonant_scattering, review_Roche, omar_spin_2015} and superior spin transport properties \cite{pep_2015, stamfer}, offers a unique platform to study the interaction of impurities with the electron spin via the universally observed phenomenon of $1/f$ noise. 
Such an approach leads to the expectation of a \emph{spin-dependent} $1/f$ noise in the average spin accumulation \cite{Tombros_nature}, and to the fundamental question of what is the origin of this noise. 
In our recent experiment, we measured for the first time the spin-dependent $1/f$ noise \cite{omar_spin_2017}. For this, we used graphene as a prototypical spin channel, leading to two major observations. First, we extracted a noise magnitude $\gamma$ for spin transport, i.e., $\gamma^{\text{s}}$ three to four orders of magnitude higher than for charge transport ($\gamma^{\text{c}}$), attributed to a drastically enhanced spin scattering as compared to charge scattering.
Second, we identified that the spin-dependent noise was dominated by the noise from the spin-relaxation processes. 

In this work, we develop a two-channel resistor model and using this, simulate the charge $1/f$ noise of similar magnitude as that experimentally measured in ref.~\cite{omar_spin_2017}, employing $\gamma=\gamma^{\text{c}}\sim$~5$\times$10$^{-8}$. Next, we use the same model to simulate the spin signal and the spin-dependent $1/f$ noise in the \emph{nonlocal} geometry. The simulated spin signal is in agreement with the experimental results. Nevertheless, we find that the simulated spin-dependent noise is significantly lower than the experimental counterpart, using the noise magnitude $\gamma\sim\gamma^{\text{c}}$ for each process. Via further analysis, we show that an agreement with the measured $1/f$ spin-dependent noise \cite{omar_spin_2017} is obtained by considering $\gamma\simeq$10$^{4}\times\gamma^{\text{c}}$, i.e.\ $\gamma \sim \gamma^{\text{s}}$, only for the spin-relaxation resistances. This leads to a quantitative demonstration of a spin-dependent noise dominated by the spin-relaxation processes with a large $\gamma$.

\begin{figure}
 \includegraphics[scale=1]{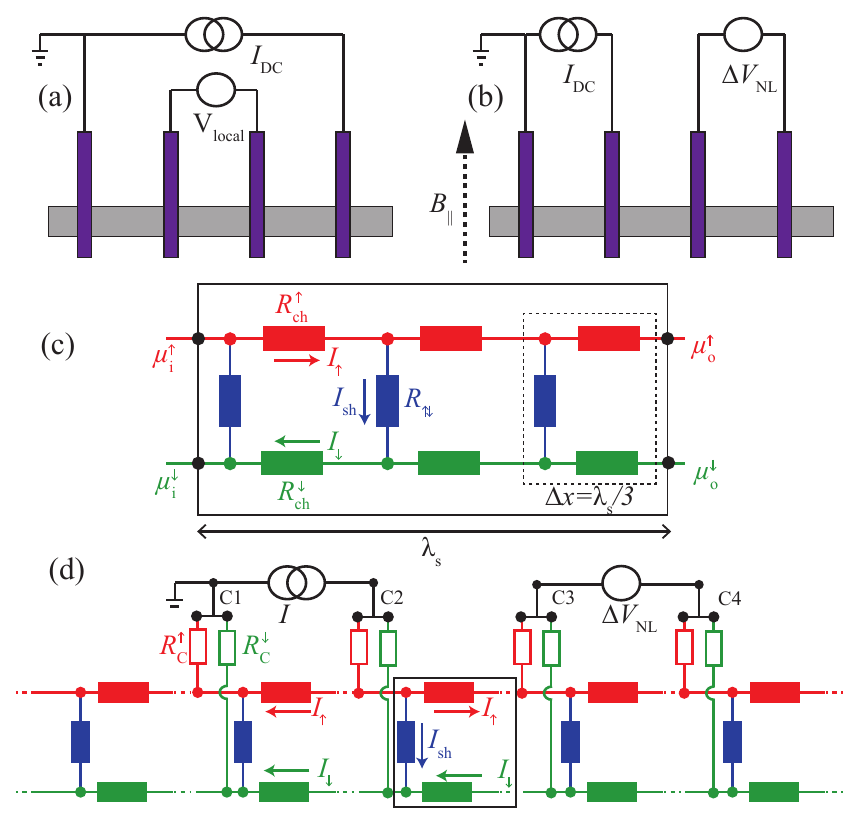}
 \caption{ \label{non local ckt} 
 Schematic diagram of nonmagnetic channel (gray) with spin-polarized contacts (purple) for a four probe (a) charge transport and (b) spin transport measurement scheme. 
A region of length $l\approx \lambda_\text{s}$ is modeled as $n$=3 basic units connected in series (c), each formed by an equivalent circuit of a spin-up resistance $R_{\text{ch}}^{\uparrow}$ (red) and a spin-down resistance $R_{\text{ch}}^{\downarrow}$(green), connected via a spin-relaxation resistance $R_{\uparrow \downarrow}$ (blue). 
(d) A two-channel model for spin transport is constructed by replacing the transport channel with a series connection of basic units from (c), and by modeling 
the spin-polarized contacts with two resistors $R_{\text{C}}^{\uparrow}$ and $R_{\text{C}}^{\downarrow}$.
{
}
}
\end{figure}

Our elementary two-channel resistor model for the four-probe nonlocal geometry (Fig.~\ref{non local ckt}(b)) \cite{Tombros_nature}, is developed as an extension to ref.~\cite{fert_conditions_2001}. A region of length $l$ in the device is modeled as $n$ basic units connected in series, with each unit corresponding to the spin transport within a length $\Delta x$. 
For our simulations, we consider $\Delta x = \lambda_{\text{s}} / 3$, with $\lambda_{\text{s}}$ the spin relaxation length in the channel, as shown in Fig.~\ref{non local ckt}(c). One channel unit is represented by a spin-up and a spin-down channel resistances, $R_{\text{ch}}^{\uparrow}$ and $R_{\text{ch}}^{\downarrow}$, connected via a spin relaxation resistance $R_{\uparrow \downarrow}$. 
The resistance to a charge current, for a channel length $\Delta x$ and width $w$, is $R_{\text{ch}}=R_{\text{sq}}\Delta{x}/w$, with $R_{\text{sq}}$ the square resistance. For a two-channel model, this can be represented as a parallel configuration of $R_{\text{ch}}^{\uparrow}$ and $R_{\text{ch}}^{\downarrow}$, both expressed as,
\begin{equation}
R_{\text{ch}}^{\uparrow}=R_{\text{ch}}^{\downarrow}=2\times R_{\text{ch}}=\frac{2R_{\text{sq}}\Delta x }{w},
\label{channel resistance}
\end{equation}
which holds true due to the non-magnetic nature of the channel. To complete the model of the channel we introduce the spin relaxation resistance $R_{\uparrow \downarrow}$ given by,
\begin{equation}
 R_{\uparrow \downarrow}=\frac{2R_{\text{sq}}\lambda_{\text{s}}^2}{w\Delta x},
 \label{relax resistance}
\end{equation}
which corresponds to the spin relaxation within a channel length of $\Delta x$ (see Supplemental Information for derivation).
Within the transport channel there are two current branches,  $I_{\uparrow}$ (upper branch) and $I_{\downarrow}$ (lower branch), see Fig.~\ref{non local ckt}(d). In the nonlocal part of the circuit, where the charge current is zero, $I_{\text{c}}=I_{\uparrow}+I_{\downarrow}=0$ , there exists only a pure spin current, $I_{\text{s}}=I_{\uparrow}-I_{\downarrow}\neq 0$.
Therefore, the spin accumulation, $\mu_{\text{s}}$, i.e., the difference between the the chemical potentials in upper branch, $\mu_{\uparrow}$, and the lower branch, $\mu_{\downarrow}$, is present only  due to spin transport in the channel.

With respect to the contacts, each spin-polarized injector (detector) is represented as a combination of two resistors, 
$R_{\text{C}}^{\uparrow}$ and $R_{\text{C}}^{\downarrow}$, 
corresponding to injection into the spin-up and spin-down channels, as shown in Fig.~\ref{non local ckt}(d). These resistors must satisfy the following conditions \cite{maassen_contacts} regarding the measured contact polarization, $P$, and the measured contact resistance, $R_\text{C}$, namely,
\begin{equation}
 P=\frac{R_{\text{C}}^{\downarrow}-R_{\text{C}}^{\uparrow}}{R_{\text{C}}^{\downarrow}+R_{\text{C}}^{\uparrow}},
\text{ and }
 R_{\text{C}}=\frac{R_{\text{C}}^{\downarrow}R_{\text{C}}^{\uparrow}}{R_{\text{C}}^{\downarrow}+R_{\text{C}}^{\uparrow}},
 \label{contact resistance}
\end{equation}
to achieve consistency between the experiment and the modelling. For the simulation, we use $R_{\text{C}}^{\uparrow}(R_{\text{C}}^{\downarrow})\sim$ 9.5 k$\Omega$(10.5 k$\Omega$), i.e. , corresponding to $P\sim$5\% and $R_{\text{C}} \sim$5 k$\Omega$, $R_{\text{ch}}^{\uparrow}=R_{\text{ch}}^{\downarrow}$=200 $\Omega$, and $R_{\uparrow\downarrow}$=4.7 k$\Omega$.
The nonlocal spin signal $\Delta V_{\text{NL}}$ due to an injection current $I$, can be estimated using \cite{popinciuc_electronic_2009, maassen_contacts},
\begin{equation}
 \Delta V_{\text{NL}}= \frac{P^2 I R_{\text{sq}}\lambda_{\text{s}}e^{-L/\lambda_{\text{s}}}}{2w},
 \label{drnl}
\end{equation}
where $L$ is the separation between the injector and detector contacts. For the device used in ref.~\cite{omar_spin_2017}, $L \sim \lambda_{\text{s}}$.

\begin{figure}
 \includegraphics[]{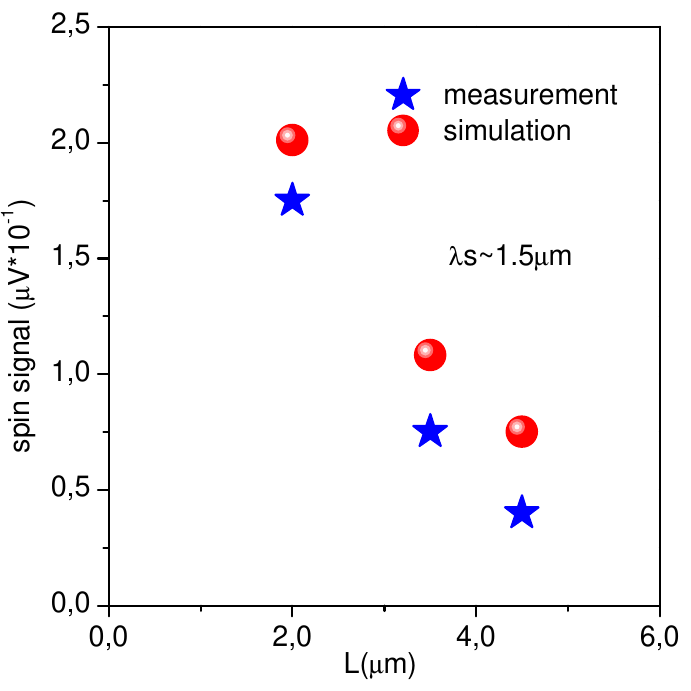}
 \caption{\label{spin transport measured simulated} Spin signal as a function of injector-detector separation $L$. 
 For the circuit simulations, we use the experimentally obtained $P \sim$ 5\%, $\lambda_{\text{s}} \sim 1.5~$~\textmu m, $R_{\text{sq}} \sim 400~\Omega$ and $w = 1.7$~\textmu m.}
\end{figure}

All values for the parameters in Eq.~\ref{drnl} are experimentally obtained and consequently used to construct the two-channel model shown in Fig.~\ref{non local ckt}(d), by using Eqs.~\ref{channel resistance}--\ref{contact resistance}. 
In order to check the validity of our model for spin transport, we compare the simulated spin signal with the experimental values \cite{omar_spin_2017}. First, we consider the measured spin signal for the graphene spintronic device at different values of $L$. Next, we apply our circuit model from Fig.~\ref{non local ckt}(d), using the corresponding number of repetitions for our basic unit (Fig.~\ref{non local ckt}(c)), therefore replicating the experimental device for the same values of $L$. The agreement between the experiment and the calculated spin signal, shown in Fig.~\ref{spin transport measured simulated}, confirms the validity of our model.     

Let us now consider electronic noise in our model. At equilibrium, in the absence of any charge current, there is always a finite thermal noise present in a transport channel. However, in a non-equilibrium situation due to a charge current $I$, a frequency dependent $1/f$ noise is present and dominates at low frequencies. This noise is generated due to the trapping-detrapping of charge carriers at a finite time scale, via impurities present at the contact-channel interface or at the substrate-transport channel interface \cite{dutta_low-frequency_1981, pal_microscopic_2011}. For the case of spin transport, a spin-dependent $1/f$ noise can be generated either by fluctuations in contact polarization (during spin injection/detection) \cite{jiang_low-frequency_2004,ingvarsson_electronic_1999}, or by fluctuations in channel or spin relaxation resistors (during spin transport). 
The observed $1/f$ spin-dependent noise is believed to originate from the spin relaxation processes \cite{omar_spin_2017}.

\begin{figure}[tbp]
 \includegraphics[scale=1]{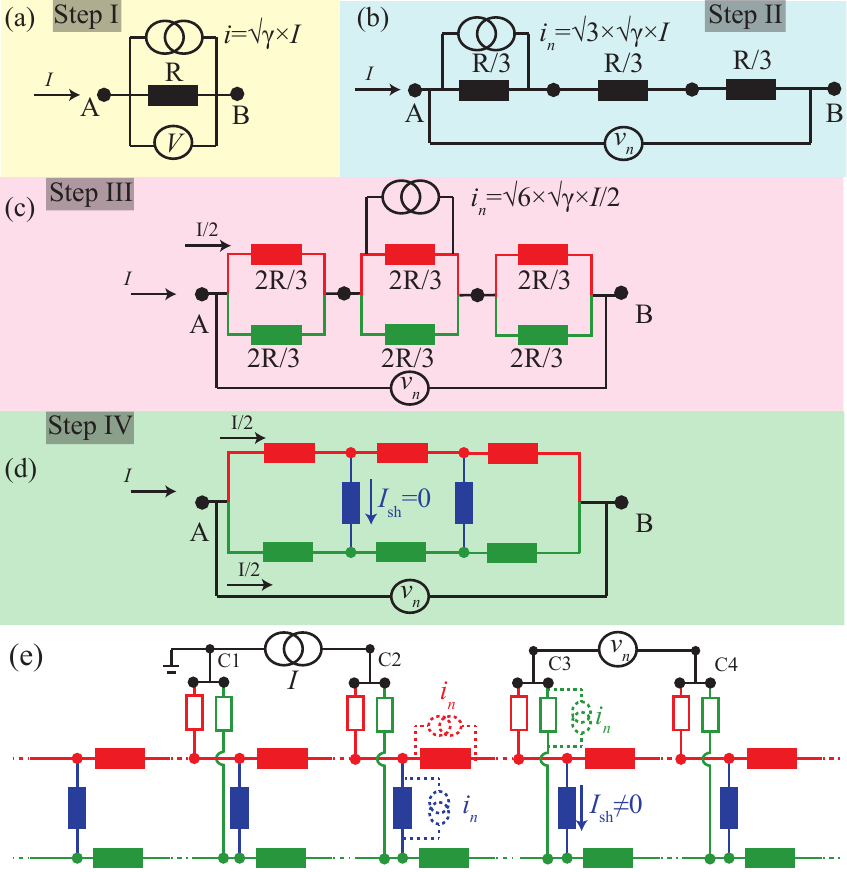}
 \caption{\label{local-nl-circuit} Circuit model for $1/f$ noise (at 1~Hz) for (a) a single-channel region of resistance $R$, under an applied current $I$, with a current noise source $i = \sqrt{\gamma} I$, has total noise $V$. (b) Same region as in (a), represented as a series of three resistors, each with a scaled $i_n = \sqrt{3 \gamma}  I$, which keeps  the total noise consistent, $V$ \cite{razavi_fundamentals_2012}, using Eq.~\ref{vtotal}.
(c) Transition to a two-channel model in the limit of fast spin relaxation, where the channel resistors, $R_{\text{ch}}^{\uparrow (\downarrow)}=2R/3$, each have an equivalent noise current source $i_n = \sqrt{6 \gamma} \times I/2$. 
(d) Introduction of the spin relaxation resistance, $R_{\uparrow \downarrow}$. For an unpolarized current $I$, there is no current present at these resistors, so they do not contribute to $1/f$ noise. To keep a consistent $V$ for cases (a)--(d), we must consider $i_n$ in the spin channel resistors as independent noise sources. 
(e) Full two-channel model, as in Fig.~\ref{non local ckt}(d), 
including also noise sources for the contact resistances and for the spin relaxation resistances. The latter contribute to the total noise, as in a spin injection geometry there is now a spin current present in the channel. A noise voltage $v_n$ appears between C3-C4 due to noise current $i_n$ in the circuit}
\end{figure}

In the present work, we simulate the charge and the spin-dependent noise originating from the contacts, the channel, and the spin-relaxation resistances, and analyze their individual contributions to find out the dominant source of spin-dependent noise. 
Noise associated with each of these resistor elements is represented as a root mean squared (\emph{rms}) current noise source, $i$, in parallel with the noiseless resistor $R$, as shown in Fig.~\ref{local-nl-circuit}(a). For a noise spectral density $S$~[A$^2$Hz$^{-1}$] at the element $R$, the equivalent noise current is $i \equiv S$$^{1/2}$~[A~Hz$^{-1/2}$]. 
For each noise source $i_n$ applied across a resistor $R_n$, we must evaluate the corresponding noise voltage appearing between the detector contacts, $v_n = \eta_n i_n$. 
Here, $\eta_n$ is a coefficient that depends on the circuit topology, relating the element to the measurement contacts, and therefore depends on the measurement geometry. The total noise, $V$, due to all circuit elements will be, 
  \begin{equation}
		V=\sqrt{v_1^2+v_2^2+v_3^2+\ldots+v_{n-1}^2+v_n^2},
		\label{vtotal}
  \end{equation}
where we assume that all noise sources are independent. This condition is necessary to achieve a consistent description of the total noise, $V$, as shown in Fig.~\ref{local-nl-circuit}(a)--(e) and in the Supplemental Section.

\begin{figure}
 \includegraphics[]{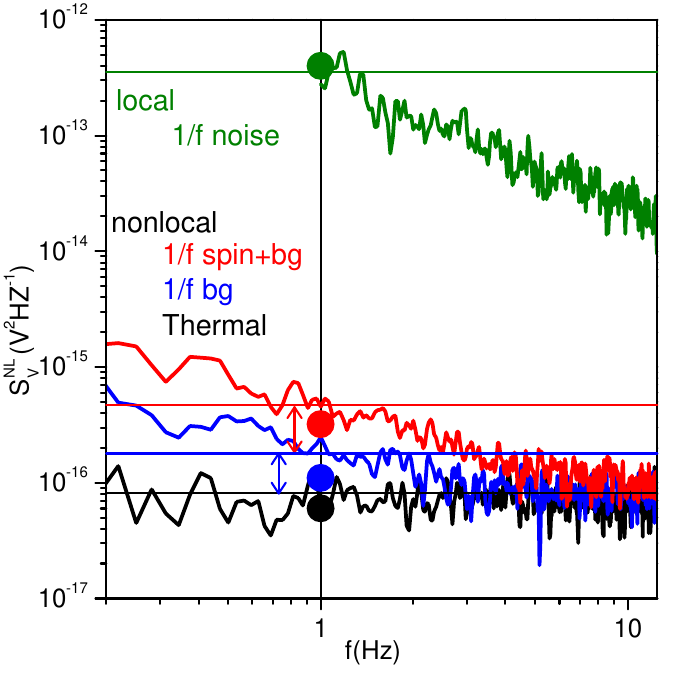}
 \caption{\label{spin noise simulation} Charge and spin-dependent $1/f$ noise measurements for the experimental device of Ref.~\cite{omar_spin_2017}. 
The green spectrum is the charge $1/f$ noise measured in the local four-probe geometry. 
The black spectrum is the nonlocal thermal noise background, for $I=0$ in Fig.~\ref{local-nl-circuit}(e). The red spectrum is the measured total nonlocal $1/f$ noise, which is the sum of the thermal noise, the background charge noise, and the spin-dependent noise (magnitude denoted by the red arrow). The blue spectrum is the spin-independent background (thermal and charge backgrounds), measured when the spin accumulation is suppressed by an applied out of plane magnetic field.  The horizontal lines indicate the noise levels measured at 1~Hz, and the dots the corresponding results from the simulations.
}
\end{figure}

We start by calculating thermal noise between C3-C4, using the circuit model of Fig.~\ref{local-nl-circuit}(e) with $I=0$,  as a test case for our model. This contribution acts as a background noise at C3-C4, which we simulate for each element $R_n$, by considering a current noise spectral density $S_n = 4k_{\text{B}}T \Delta f / R_n$. The resulting equivalent noise current source $i_n$ across each element, is then used to calculate the total noise voltage at $\Delta f=1$~Hz for the nonlocal measurement geometry, according to Eq.~\ref{vtotal}. In this way, we can estimate the contribution from the spin-relaxation resistors, channel resistors, and contacts, separately (1$^{\text{st}}$ column of Table.~\ref{tab:spin noise summary}).
The simulation result for the nonlocal thermal background, $\sim 6 \times 10^{-17}$~V$^2$~Hz$^{-1}$ (see Table~\ref{tab:spin noise summary}), is in good agreement with the measured thermal noise, $\sim 10^{-16}$~V$^2$~Hz$^{-1}$, as shown in Fig.~\ref{spin noise simulation},  supporting the validity of the model also for the noise simulations.

Next we proceed to consider $1/f$ noise, first in the local measurement configuration. For a local measurement as in Fig.~\ref{non local ckt}(a), the (charge) $1/f$ noise spectral density $S_{\text{I}}^{\text{local}}$ has a frequency and current dependent power spectral density, described by the Hooge formula,
\begin{equation}
	\label{hooge charge}
	S_{{I}}^{\text{local}}=\frac{\gamma^{\text{c}}I^2}{f^{\alpha}},
\end{equation}
where $\alpha \sim 1$ and $\gamma^{\text{c}}$ is the charge noise magnitude. The latter is defined as the Hooge parameter, $\gamma_{\text{H}}^{\text{c}}$, divided by the total number of carriers in the transport channel, i.e.\ $\gamma^{\text{c}}=\gamma_{\text{H}}^{\text{c}}/(n w L)$, where $n$ is the (2D) charge carrier density. 
From our measurements of a graphene device we obtained $\gamma^\text{c} \sim 5\times10^{-8}$ at $f=1$~Hz \cite{omar_spin_2017}, of a similar magnitude as in ref.\cite{pal_microscopic_2011, balandin_low-frequency_2013}. 

For our calculations of $1/f$ noise we consider this value of $\gamma^\text{c}$. We first proceed to scale the experimental noise magnitude $\gamma^\text{c}$ with respect to the length of the basic unit element in our two-channel model, ($\Delta x =\lambda_{\text{s}}/3\sim L/3$), as shown in Fig.~\ref{local-nl-circuit}(a)--(c). This results in $\gamma_{\text{scaled}} = 6 \gamma^\text{c}$ for the spin channel resistors. Each resistor element has an equivalent current noise source $i_n = \sqrt{S_{\text{I}}}= \sqrt{\gamma_{\text{scaled}}} I_n=\sqrt{6\gamma^{\text{c}}} I_n$ for f~=~1 Hz. For the contacts, $\gamma = \gamma^{\text{c}}_{\text{contact}} \sim 2\times 10^{-8}$ is used for the calculation, as obtained experimentally by measuring the $1/f$ noise across the contacts \cite{omar_spin_2017}. Furthermore, we must calculate the current $I_n$ through each resistor element, for the specific measurement geometry under consideration. For the local configuration we consider an applied dc current $I = 10$~\textmu A between contacts C1--C4, similar to the experiment. In this way we can obtain the equivalent noise current sources, $i_n$, for all the elements, and subsequently calculate their contribution to the total noise at the detector contacts C2--C3, using Eq.~\ref{vtotal}. Here it is relevant to clarify the role of the spin-relaxation resistors, $R_{\uparrow \downarrow}$, and the corresponding noise magnitude $\gamma_{\uparrow \downarrow}$. In the local geometry of Fig.~\ref{non local ckt}(a), we do not expect to inject any significant spin accumulation within the center of the channel, using the similar circuit of Fig.~\ref{local-nl-circuit}(e). Here, we assume that the outer contacts are situated far away, which results in negligible spin-accumulation between the detector electrodes C2-C3. 
As an initial estimation, we assume that the noise from the charge scattering and spin-relaxation have same origin and use $\gamma_{\uparrow \downarrow} =3\times \gamma^\text{c}{(\Delta x \sim L/3)}$ (see supplemental Information for details). 
The simulation results for the local measurement show that the contribution towards $1/f$ charge noise from the spin-relaxation resistors is $\sim { 10^{-20}}$~V$^2$~Hz$^{-1}$, which is seven orders of magnitude lower than the experimentally obtained noise magnitude of $\sim 3 \times 10^{-13}$~V$^2$~Hz$^{-1}$. On the contrary, the calculated contribution from the channel spin resistors, $R_\text{ch}^{\uparrow (\downarrow)}$ amounts to a noise of $\sim {4\times 10^{-13}}$~V$^2$~Hz$^{-1}$, implying that the charge 1/f noise is  dominated by the noise from the channel resistors with $\gamma \sim \gamma^{\text{c}}$, as shown in Fig.~\ref{spin noise simulation}. The 1/f scaling of the calculated noise is straightforward. It can be obtained at any frequency $f$ by replacing each noise current $i_n$ with $i_n/\sqrt{f}$ and recalculating the output noise with the modified $i_n$. Alternatively, it can be shown using Eq.~\ref{vtotal} that the total noise power $S^{1/f}\propto 1/f$.  Therefore, the noise simulated at one frequency can be scaled with the factor 1/f to obtain the frequency dependent behavior.

Finally, we consider the nonlocal noise. In analogy to Eq.~\ref{hooge charge}, the spin-dependent contribution to the $1/f$ noise, $\Delta S_{{V}}^{\text{NL}}$, can be expressed as,
 \begin{equation}
  \Delta S_{{V}}^{\text{NL}}=\frac{\gamma^{\text{s}} \Delta {V}_{\text{NL}}^2}{f^{a}},
  \label{eq:Hooge spin}
 \end{equation}
where $\gamma^{\text{s}}=\gamma_{\text{H}}^{\text{s}}/(n w \lambda_{\text{s}})$ is the noise magnitude for spin transport, and $\gamma_{\text{H}}^{\text{s}}$ represents the Hooge parameter for spin transport. Here, we consider the non-conserved nature of the spin current and take $\lambda_{\text{s}}$ as the characteristic length for the normalization of $\gamma_{\text{H}}^{\text{s}}$ by the total number of carriers in the channel under measurement. $\Delta V_{\text{NL}}= P\mu_{\text{s}}/e$ is the measured nonlocal spin signal due to the average spin accumulation $\mu_{\text{s}}$.

 We use the nonlocal measurement configuration, Fig.~\ref{local-nl-circuit}(e) for simulating the spin-dependent $1/f$  in order to eliminate the contribution of charge noise. We consider a dc current $I = 10$~\textmu A between contacts C1--C2 and calculate the noise between C3--C4 due to each circuit element, following the procedure established for the thermal and charge $1/f$ noise simulations, assuming the same $\gamma_{\text{scaled}} \sim 6\gamma^{\text{c}}$ for $R_{\text{ch}}^{\uparrow},R_{\text{ch}}^{\downarrow}$, $\gamma_{\uparrow \downarrow}\sim 3\gamma^{\text{c}}$ for $R_{\uparrow \downarrow}$, and $\sim {2\times10^{-8}}$ for the spin-polarized contacts. Here, we assume that the mechanisms, producing the $1/f$ charge and spin-dependent noise are same. The simulated $1/f$ noise results in $\sim {5\times 10^{-17}}$~V$^2$~Hz$^{-1}$,  which is lower by an order than the experimental counterpart, $\sim 5 \times 10^{-16}$~V$^2$~Hz$^{-1}$, i.e. the magnitude denoted by the red arrow in Fig.~\ref{spin noise simulation}. In particular, the noise from the spin-relaxation resistances, which was identified as a dominant noise source in the measurements \cite{omar_spin_2017}, is only $\sim  10^{-20}$~V$^2$~Hz$^{-1}$, so lower than the measured spin-dependent noise by almost four orders of magnitude. 
From the simulation results it is clear that the processes producing the spin-dependent $1/f$ noise are very distinct from that of charge $1/f$ noise and cannot be explained by the noise magnitude $\gamma^{\text{c}}$ associated to the charge $1/f$ noise. 
In fact, the calculated nonlocal $1/f$ noise is in a better agreement with the measured nonlocal $1/f$ spin-independent background noise, given by the magnitude of the blue arrow in Fig.~\ref{spin noise simulation}. An agreement with this background, present when an out-of-plain magnetic field is applied and there is no spin accumulation present at the detector due to dephased spins, suggests that with the present consideration we only capture the nonlocal contribution from the noise sources in the local circuit, where a current is present, but not that contribution originating from the nonlocal spin transport. Given that the noise sources of the channel resistances and the contacts are experimentally determined, the only unknown noise sources are those related to the spin relaxation resistors, which up to now we have considered to be $\gamma_{\uparrow \downarrow}$= 3$\times \gamma^\text{c}$.

\begin{table}
 \vspace{0.5 cm}
\begin{ruledtabular}
\begin{tabular}{ccccccc|cccc}

\multicolumn{7}{c|}{thermal noise (V$^2$~Hz$^{-1}$)}&
\multicolumn{3}{c}{$1/f$ noise (V$^2$~Hz$^{-1}$)}\\
\hline

\multicolumn{6}{c}{}&{}&
\multicolumn{2}{c}{$\gamma^{\text{s}}$=5$\times$10$^{-8}$}&
\multicolumn{1}{|c}{$\gamma^{\text{s}}$=5$\times$10$^{-4}$}\\
\hline
 \multicolumn{6}{c|}{$R_{\text{C}}^{\uparrow (\downarrow)}$}&{5$\times$ 10$^{-17}$}&
\multicolumn{2}{c}{3.2$\times$10$^{-17}$}\\
\hline
\multicolumn{6}{c|}{$R_{\text{ch}}^{\uparrow (\downarrow)}$}&{9$\times$10$^{-18}$}&
\multicolumn{2}{c}{1.5$\times$10$^{-17}$}&

\\
\hline
\multicolumn{6}{c|}{$R_{\uparrow \downarrow}$}&{10$^{-21}$}&
\multicolumn{2}{c}{2$\times$10$^{-20}$}&
\multicolumn{1}{|c}{2$\times$10$^{-16}$}
\\
\hline
\hline
\multicolumn{6}{c|}{total}&{6$\times$10$^{-17}$}&
\multicolumn{2}{c}{4.7$\times$10$^{-17}$}&
\multicolumn{1}{|c}{3.1$\times$10$^{-16}$}
\\
 \end{tabular}
\end{ruledtabular}
\caption{\label{tab:spin noise summary} Summary of thermal noise and $1/f$ nonlocal noise contributions from injection/detection contacts, transport spin resistors (channel), and spin-relaxation resistors (spin-flip processes), obtained from simulations with a two-channel model.}

\end{table}

From the spin-dependent noise measurements in ref.~\cite{omar_spin_2017}, we experimentally obtained the spin noise magnitude $\gamma^{\text{s}} \sim 10^{-4}$ -- $10^{-3}$, by fitting the dependence of $\Delta S_{\text{V}}^{\text{NL}}$ on the spin signal $\Delta V_{\text{NL}}$ with Eq.~\ref{eq:Hooge spin}. 
This value was surprisingly up to four orders of magnitude higher than $\gamma^{\text{c}}$ for the charge noise. 
The main question is to find out to which process we can assign this $\gamma^{\text{s}}$, which would result in a simulated total $1/f$ noise closer to the experimental value. 
Let us briefly consider the case where we use this experimental $\gamma^{\text{s}}$ to calculate the noise from the channel and contact resistors. This exercise results in a $1/f$ noise level $\sim 10^{-13}$~V$^2$~Hz$^{-1}$,  higher by three orders of magnitude than the observed noise level in the experiments. This result indicates that, according to our circuit model, the experimental $\gamma^{\text{s}}$ can not be assigned to the channel nor the contact resistances. 
Therefore, we forgo our initial consideration of $\gamma_{\uparrow \downarrow}$= 3$\times \gamma^\text{c}$, and recalculate the nonlocal $1/f$ noise for the case of a spin-relaxation resistance noise magnitude given by the experimentally measured spin noise magnitude, i.e.\ $\gamma_{\uparrow \downarrow}= 3\times\gamma^\text{s}$. The results shown in the rightmost column of Table~\ref{tab:spin noise summary}, demonstrate a similar magnitude for the $1/f$ nonlocal noise due to the spin-relaxation resistors, $\sim  {2\times10^{-16}}$~V$^2$~Hz$^{-1}$, to the measured spin-dependent $1/f$ noise in \cite{omar_spin_2017}, shown by the red arrow in Fig.~\ref{spin noise simulation}. Based on simulation results, we argue that $\gamma_{\uparrow \downarrow}$ is orders of magnitude higher than $\gamma$ of the channel resistors. More importantly, it is in a quantitative agreement with the experimentally obtained $\gamma^{\text{s}}$.  

In conclusion, we present a two-channel model to simulate $1/f$ noise, associated with both charge and spin transport. The noise contribution from different circuit elements demonstrates that the measured spin-dependent $1/f$ noise in Ref.~\cite{omar_spin_2017} is dominated  by the noise from the spin-relaxation resistances, in quantitative agreement with spin relaxation processes with noise magnitude $\gamma_{\uparrow \downarrow} \simeq 10^{3-4}\times \gamma^{\text{c}}$. Our approach provides a simplified platform to understand and address the complex nature of the noise related to spin-transport experiments and enables the proposal of noise measurements as a direct tool to probe the nature of spin-relaxation.

\begin{acknowledgments}
This research work was financed under EU-graphene flagship program Core-I project (190637100) and supported by the Zernike Institute for Advanced Materials, the Netherlands Organization for Scientific Research (NWO) and the Future and Emerging Technologies (FET) programme within the Seventh Framework Programme  for Research of the European Commission, under FET-open Grant No. 618083 (CNTQC).
\end{acknowledgments}

%

 \newpage
 
 \beginsupplement
 \begin{center}
  \textbf{\large Supplementary Information}
 \end{center}

 \section{Derivation for spin relaxation resistance}
 
 The expression for the spin-relaxation resistance in the circuit can be derived easily. The current $I_{\text{sh}}$ corresponds to the spin relaxation within the volume $w\Delta x$, represented by the  relaxation resistance $R_{\uparrow \downarrow}$ is given by:
 
 \begin{equation}
  I_{\text{sh}}= \frac{V_{\uparrow}-V_{\downarrow}}{R_{\uparrow \downarrow}}=\frac{\mu_{\uparrow}-\mu_{\downarrow}}{e(R_{\uparrow \downarrow})}
  \label{relax1}
 \end{equation}
 
 \begin{equation}
  \frac{I_{\text{sh}}}{w\Delta x}=\frac{e(n_{\uparrow}-n_{\downarrow})}{\tau_{\text{s}}}=\frac{eN(\mu_{\uparrow}-\mu_{\downarrow})}{\tau_{\text{s}}}
  \label{relax2}
 \end{equation}
 
 Here, $n_{\uparrow}(n_{\downarrow})$ is the number of spin up(down) electrons, $\tau_{\text{s}}$ is the spin-flip time, and $N$ is the electron density of states at Fermi energy. The expression can be simplified using the Einstein relation:
 \begin{equation}
  \frac{1}{R_{\text{sq}}}=Ne^2D
  \label{Einstein}
 \end{equation}
 where $R_{\text{sq}}$ is the sheet resistance of the channel and $D$ is the diffusion coefficient. Replacing Eq.\ref{Einstein} and $D={\lambda_{\text{s}}}^2/\tau_{\text{s}}$ into  Eq.\ref{relax2}, we can rewrite $I_{\text{sh}}$ as
 
 \begin{equation}
 I_{\text{sh}}=\frac{\mu_{\uparrow}-\mu_{\downarrow}}{eR_{\text{sq}}{\lambda_{\text{s}}}^2}
 \label{relax3}
 \end{equation}
 
 By solving Eq.\ref{relax2} and Eq.\ref{relax3}, we obtain the expression for $R_{\uparrow \downarrow}$
 
 \begin{equation}
  R_{\uparrow \downarrow}=\frac{2R_{\text{sq}}\lambda_{\text{s}}^2}{w\Delta x}
 \end{equation}

 \section{scaling of noise current in a two channel model}
 
 In Eq.~\ref{hooge charge} 
 \begin{equation}
  \gamma^{\text{c}} =\frac{\gamma_H^{\text{c}}}{n\times W \times L}
  \label{gamma charge}
 \end{equation}
 and
 \ref{eq:Hooge spin} of the main text:
 \begin{equation}
  \gamma^{\text{s}}=\frac{\gamma_H^{\text{c}}}{n\times W \times \lambda_{\text{s}}}
  \label{gamma spin}
 \end{equation}
 
   the noise magnitudes of the charge (spin) transport channel $\gamma^{\text{c}}(\gamma^{\text{s}})$ need to be scaled with respect to the carrier concentration and device parameters ($W,L$), in oder to estimate the accurate noise current $i_n$ of the resistor, unlike in the case of the intrinsic $\gamma_H^{\text{c}} (\gamma_H^{\text{s}})$, which are constant. Note that $L$, here is the separation between the inner injector and detector electrodes, i.e., the transport channel.
 
 In this section, we build a two channel model and explain the scaling of $\gamma$, that can be either $\gamma^{\text{c}}$ or $\gamma^{\text{s}}$ associated with the resistors, that converts the charge or spin current to a noise current. 
 \begin{figure}
  \includegraphics[]{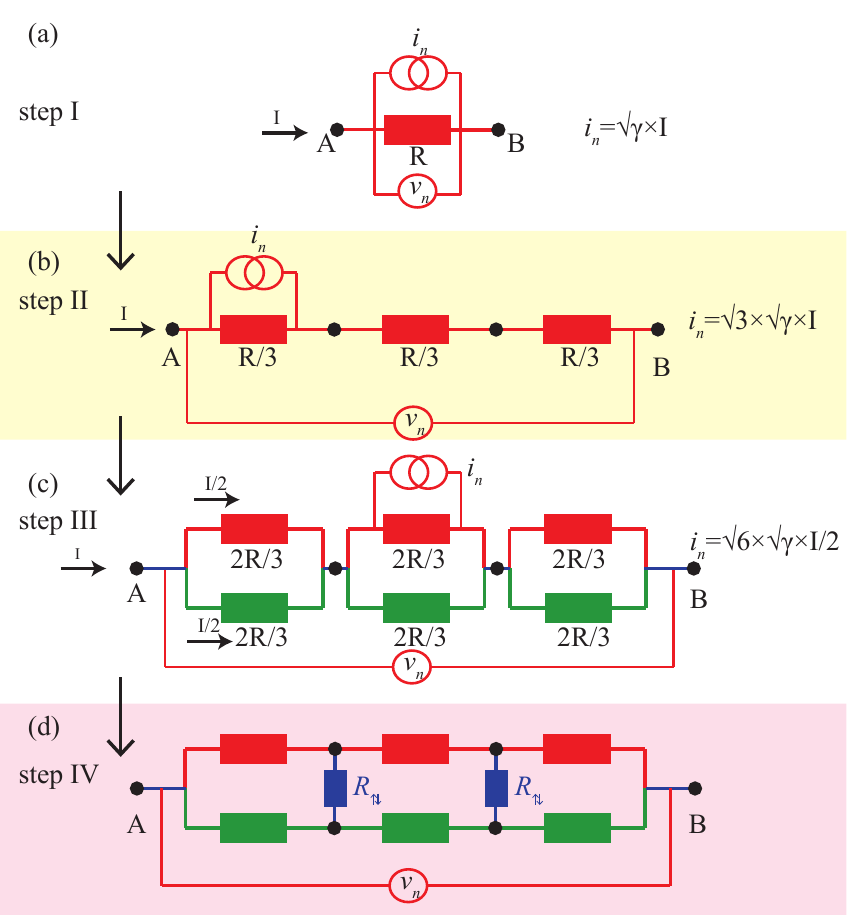}
  \caption{\label{equivalent noise}(a) noise representation of a resistor of resistance $R$. (b) in a series and (b) parallel arrangements. (d) A two channel model for spin-transport channel. The spin relaxation resistances $R_{\uparrow \downarrow}$ are shown in blue. $R_{\uparrow \downarrow}$ are inactive in the absence of a spin-accumulation in the transport channel due to symmetry of the circuit.}
 \end{figure}
 In Fig.~\ref{equivalent noise}(a), a charge(spin) current $I$ is flowing in a resistance $R$ of length $l=L\sim\lambda_{\text{s}}$ and width $W$.  This current produces a $1/f$ noise current $i_n~=~\sqrt{\gamma}I$, which is measured as noise voltage $v_n$ due to $i_n$ flowing in $R$, i.e.,
 \begin{equation}
  v_n=i_n \times R= \sqrt{\gamma} IR  
  \label{first}
 \end{equation}
 
 Now, in step II (Fig.~\ref{equivalent noise}(b)), the same $R$ is represented in a series of three resistances of $R/3$. The total resistance still remains $R$. However, the noise and the noise magnitude associated with each $R/3$ resistance is changed. Since, the length of the transport channel for each resistance is  $L/3$, using Eq.~\ref{gamma charge}, we obtain, $\gamma_{\text{new}}=3\times \gamma$. Now the noise voltage  between A and B  due to $R/3$ is:
 \begin{equation}
  v_n=\sqrt{3 \gamma} \times I \times \frac{R}{3} 
  \label{series}
 \end{equation}
  All $R/3$ resistances produce the equal amount of noise $v_n$ and the total noise $V$, assuming that all noise sources are independent and using Eq.~\ref{vtotal}, will be $\sqrt{3\times v_n^2}$, which is same as the noise measured in case of a single resistance of resistance $R$. 
  
  In step III (Fig.~\ref{equivalent noise}(c)), we divide the A-B branch into two parallel paths, and represent the equivalent resistance in a two-channel model, which is later used to model the spin transport. The net resistance still remains the same ($R_{\text{eq}}=R$). However, each $R/3$ in Fig.~\ref{equivalent noise}(b) is represented as a parallel combination of two $2R/3$ resistances. Now, for each $2R/3$ resistance the channel length and width are $L/3$ and $W/2$, respectively, due to which $\gamma_{\text{new}}=6\times \gamma$, using Eq.~\ref{gamma charge}. It should be noted that the charge current through each $2R/3$ is $I/2$, and $i_n$, therefore, will be =$\sqrt{\gamma_{\text{new}}} \times I/2=\sqrt{3/2 \times \gamma} I$, which will flow across $2R/3||2R/3=R/3$ and produce the noise voltage $v_n=\sqrt{\frac{3}{2} \times \gamma} I \times \frac{R}{3} =\sqrt{\frac{1}{6}\times \gamma} IR$. Again, using Eq.~\ref{vtotal}, one gets $V=\sqrt{6\times v_n^2}=\sqrt{\gamma} IR$.
  
  Via this exercise, we show that by dividing a resistance $R$ into a combination of several series and parallel component resistances does not change the total noise, though the noise current $i_n$ associated with each resistance needs to be rescaled according to the new geometry. For the circuit, we simulate in the main text, we extracted the $\gamma^{\text{c}}$ experimentally for a length $L=1.5 \mu$m. For the circuit simulation, we represented the length $l\sim L$ in three segments of $L/3$, connected in series, for which we can use $\gamma_{\text{new}}=6\times \gamma$ in a two channel model, as explained above. This analysis is also valid for spin current.
  \begin{figure}
   \includegraphics[]{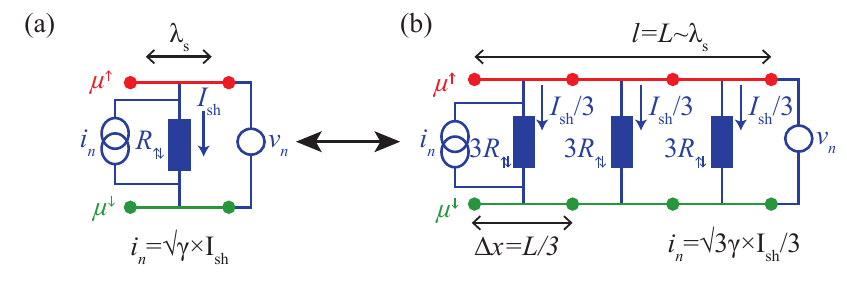}
   \caption{\label{relaxation equivalent} (a) Noise circuit representation of a spin relaxation resistance $R_{\uparrow \downarrow}$ with the current $I_{\text{sh}}$, flowing through it.  (b)Scaling of $\gamma$ for $R_{\uparrow \downarrow}$. For a homogeneous spin accumulation, i.e., $\mu^{\uparrow}=-\mu^{\downarrow}=constant$, a spin relaxation current $I_{sh}$/3 flows through each $\Delta x=\lambda_{\text{s}}$/3 unit, and $\gamma_{new}=3\times \gamma$ because of $l/\Delta x$=3.}
  \end{figure}

  Charge and spin transport in the absence of spin relaxation ($R_{\uparrow \downarrow} \sim \infty $) can be represented via Fig.~\ref{equivalent noise}(c). In the presence of spin-relaxation, which is the case for real spintronic devices, a spin relaxation resistance $R_{\uparrow \downarrow}$ (blue rectangle in Fig.~\ref{equivalent noise}(d)) is  placed as a shunt resistance between $R_{\text{ch}}^{\uparrow}$ (red rectangle) and $R_{\text{ch}}^{\downarrow}$ (green rectangle) in Fig.~\ref{equivalent noise}(d). In ref.~\cite{omar_spin_2017}, we extract $\gamma^{\text{s}}$ for the length $L \sim \lambda_{\text{s}}$, which requires the rescaling of the experimental $\gamma$ we assign to $R_{\uparrow \downarrow}$. In our circuit model, we incorporate three $R_{\uparrow \downarrow}$ resistors in the $L \sim\lambda_{\text{s}}$ scale, therefore the  $\gamma^{\uparrow \downarrow}=3 \times \gamma^{\text{c(s)}}$, is used in the simulations, and the equivalence of the noise for both the circuits in Fig.~\ref{relaxation equivalent} can be verified by the analysis presented for channel resistances.   
  
  After, we can successfully simulate the noise from the channel, it remains to figure out the unknown $\gamma$ which one should assign to calculate the noise from $R_{\uparrow \downarrow}$, which is obtained with the help of the experimental data.


\end{document}